\documentclass[12pt]{elsarticle}

\usepackage{amssymb}
\usepackage{amsmath}
\usepackage{booktabs}
\usepackage{epstopdf}
\usepackage{array}
\usepackage[version=4]{mhchem}
\usepackage{xcolor}
\usepackage{longtable}
\newcolumntype{P}[1]{>{\centering\arraybackslash}p{#1}}

\newcommand{\ms}{\phantom{-}}

\journal{JQSRT}

\begin{document}

\begin{frontmatter}

\title{High-temperature simulation of the Raman spectra of the isotopologues $^{13}$C$^{16}$O$_2$ and $^{16}$O$^{13}$C$^{18}$O}
%High-temperature Raman spectra of CO$_2$ isotopologues: simulation and comparison with experiment

%High-temperature Raman spectra of CO$_2$ isotopologues: simulation of transition intensities

\author[icn]{E. Su\'arez}
\author[FIG]{C. Ostertag-Henning}
\author[UHU,IUC]{M. Carvajal}
\author[icn]{R. Lemus\corref{cor1}}

\cortext[cor1]{Corresponding author}
\ead{renato@nucleares.unam.mx}

\affiliation[icn]{organization={Instituto de Ciencias Nucleares, Universidad Nacional Aut\'onoma de M\'exico},
city={Ciudad de M\'exico},
postcode={04510},
country={Mexico}}

\affiliation[FIG]{organization={Federal Institute for Geosciences and Natural Resources},
city={Hannover},
postcode={30655},
country={Germany}}

\affiliation[UHU]{organization={Departamento de Ciencias Integradas y Centro de Estudios Avanzados en Física, Matemáticas y Computación, Unidad Asociada GIFMAN, CSIC-UHU, Universidad de Huelva},
city={Huelva},
country={Spain}}

\affiliation[IUC]{organization={Instituto Universitario Carlos I de Física Teórica y Computacional, Universidad de Granada},
country={Spain}}

\begin{abstract}
Recently, experimental Raman spectra of the isotopologues \ce{^13C^16O2} and \ce{^16O^13C^18O} at high temperature have been reported. Separately, a reliable approach for obtaining vibrational wave functions for most isotopologues has been proposed. These descriptions, based on the $SU_1(2)\times U(3)\times SU_2(2)$ dynamical group, provide spectroscopic-quality fits to  vibrational term values with root-mean-square deviations of $0.06$ and $0.07~\mathrm{cm}^{-1}$, respectively. Using the resulting wave functions, we simulate the Raman spectra of both species by evaluating transition moments of the mean polarizability, represented as an expansion in curvilinear coordinates up to cubic terms within the same algebraic framework. \textcolor{black}{The difference of two independent experimental Raman spectra under  comparable temperature conditions is analyzed  with the help of the  simulations provided by our model.}  The resulting simulations show excellent agreement with experiment, reinforcing the reliability of the model, while the estimated transition moments are also consistent with experimental values.
\end{abstract}

%\begin{graphicalabstract}
%\includegraphics[width=\linewidth]{GraphicalAbstract}
%\end{graphicalabstract}

%\begin{highlights}
%\item Experimental vs. simulation of Raman spectra for isotopologues  \ce{^13CO2} and \ce{^16O^13C^18O}.
%\item Raman spectra between 500 and 650 K are analyzed.
%\item Diagonal ro-vibrational interactions are taken into account.
%\end{highlights}

%\begin{keyword}
%mixture of isotopologues, \ce{CO2}, molecular polarizability, transition intensities, algebraic model, Raman spectra
%\end{keyword}

\end{frontmatter}

\section{Introduction}

Because of its importance, the principal isotopologue of carbon dioxide is one of the most thoroughly studied molecules. It has received considerable attention since the early applications of IR and Raman spectroscopy \cite{herzberg}. Advances in spectroscopic techniques have led to an increase in the experimental data available for most of its isotopologues \cite{Gordon2022,Kassi20091801,Mandin1977304,Song2010332,Karlovets2014137,Karlovets201854,Toth2006243,DeGhellinck29,Perevalov2008143,Weirauch2001263,Petrova2015109,Kang20181,Li20111411,Song2011761,Borkov201457,Weirauch1999187,Karlovets201471,Campargue1999204,Karlovets201873,Yurchenko20205282,Karlovets2013116,Karlovets201489,Cermak201895,Tan201522,Elliott20141,Toth2006221,Garnache200522,Toth200743,Serdyukov2016145,Ding2004146,DeGhellinck65,Perevalov20082437,Toth2008906,Karlovets2021,Elliott201578,Ding2005117,Ding2003276}, with the exception of $^{13}$C$^{18}$O$_2$, $^{13}$C$^{17}$O$_2$, and $^{17}$O$^{13}$C$^{18}$O, for which only few vibrational terms have been reported. These new data have motivated the proposal of theoretical models capable of describing the available spectroscopic information with high accuracy \cite{Perevalov2010183,Campargue2010659,Stull19621442,Majcherova20051,Perevalov200790}. The theoretical study of the \ce{CO2} isotopologues deserves particular attention for several reasons, as demonstrated by the work of Huang \textit{et al.}~\cite{huang}, in which high-accuracy {\it ab initio} calculations were performed. One of their main applications is the determination of their natural abundance in geology \cite{Remigi2023,Cui2021,Lu2023}, which is primarily obtained through gas-phase mass spectrometry. However, this technique presents limitations due to the mass coincidence of some isotopologue fragments. This issue could be overcome by the use of IR and Raman spectroscopy. Most measurements aimed at detecting CO$_2$ are based on IR absorption and emission techniques, although these lack in spatial resolution, a feature that is provided by Raman spectroscopy \cite{Deckert}. Indeed, Raman spectroscopy represents a leading technology for probing temperature-dependent properties of gas samples.

For both IR and Raman spectroscopy, vibrational wave functions constitute the main contribution to be determined from a theoretical perspective. These wave functions can be obtained using different approaches, including perturbative treatments of the Hamiltonian, variational methods, and effective Hamiltonian models. Although these approaches differ in complexity and computational cost, in all cases the vibrational terms must be predicted with high resolution. The use of effective Hamiltonians offers the advantage of low computational cost, albeit at the expense of relying on available spectroscopic data. Furthermore, the potential energy surface is not always reliable for predictive purposes within the  Born--Oppenheimer (BO) approximation. Nevertheless, the resulting force constants can be used to assess the quality of the obtained wave functions \cite{sanchez2012}.

There are eight isotopologues of carbon dioxide with sufficient experimental energy levels to be analyzed using effective Hamiltonians, four symmetric and four asymmetric. The treatment of these isotopologues differs significantly depending on their symmetry. In the symmetric case, within a normal mode description, selection rules arise from the \textit{gerade}/\textit{ungerade} labels, whereas in the asymmetric case these rules are absent. However, the most relevant differences associated with symmetry emerge in the description based on internal coordinates. In particular, for symmetric isotopologues, the analysis is simplified by the fact that the normal coordinates coincide with symmetry-adapted coordinates. In contrast, for asymmetric isotopologues, it is necessary to employ the ${\bf GF}$ formalism in order to obtain the normal coordinates \cite{Suarezfcn, Suarezocs,Suarezn2o,Suarezn2oIso}.
 
A recent systematic study of the vibrational excitations of carbon dioxide isotopologues in their electronic ground state was carried out \cite{marisol17,marisol22,marisol24}. The model is based on the $SU_1(2) \times U(3) \times SU_2(2)$ dynamical group. The innovative aspect of this approach is the incorporation of a local mode treatment with Morse potentials associated with the stretching modes and the $U(3)$ dynamical algebra for the bending modes, while maintaining the correspondence with configuration space. A nontrivial feature of the model is that the representation of the Hamiltonian in the local scheme preserves the polyad, a pseudo-quantum number defined within a normal mode framework. This approach was used to describe the vibrational excitations of the symmetric isotopologues $^{13}$C$^{16}$O$_2$, $^{12}$C$^{18}$O$_2$, $^{12}$C$^{17}$O$_2$, and $^{12}$C$^{16}$O$_2$, as well as the asymmetric isotopologues $^{16}$O$^{13}$C$^{18}$O, $^{16}$O$^{12}$C$^{18}$O, $^{16}$O$^{12}$C$^{17}$O, and $^{17}$O$^{12}$C$^{18}$O, all of them with root-mean-square deviations around $0.10$ cm$^{-1}$. The quality of these descriptions supports the use of the corresponding vibrational wave functions for the analysis of IR and Raman spectra.

The simulation of spontaneous Raman scattering relies on the molecular polarizability tensor at certain level of approximation. For a purely vibrational description, the isotropic part provides the relevant contribution, as reflected in the trace of the scattering cross section. This contribution is expressed as a Taylor expansion in curvilinear coordinates around equilibrium. Since this expansion involves derivatives evaluated at equilibrium, a crucial step in the simulation is their determination. Due to the difficulty of obtaining them from {\it ab initio} calculations, they are instead estimated from a minimal set of reliable experimental transition moments. In this regard, \'Alvarez \textit{et al.} reported accurate transition moments for the principal isotopologue \cite{Alvarez2024}, which can be used to determine these derivatives through a fitting procedure. Once the derivatives are known and the vibrational wave functions are available, the Raman spectrum can be simulated. This procedure was applied to the principal isotopologue, comparing the results with an experimental spectrum in the range 1150--1500 cm$^{-1}$ at a temperature of 1743 K. Initial simulations assumed the polyad $P_{212}=2(\nu_1+\nu_3)+\nu_2$ \cite{Lemus2014}, while later studies explored the polyads $P_{213}=2\nu_1+\nu_2+3\nu_3$ and $P_{214}=2\nu_1+\nu_2+4\nu_3$ \cite{Bermudez2019}. An important advantage of the BO approximation is that the derivatives with respect to isotopically invariant coordinates are the same for all isotopologues, allowing the computation of transition moments once the vibrational wave functions are known. This condition is fulfilled in the present approach, where both the derivatives and the wave functions have been determined for most of the isotopologues.

In the recent contribution by \'Alvarez \textit{et al.}, experimental polarizability transition moments of the principal isotopologue CO$_2$ for excited vibrational states were obtained from Raman spectra recorded at 295, 373, 580, 950, and 1780 K \cite{Alvarez2024}. These results were used by our group to refine the corresponding polarizability derivatives, which were subsequently employed to analyze Raman spectroscopic signatures in terms of a molecular descriptor associated with the degree of normality \cite{suarez2026}. More recently, again the group of J.M. Fern\'andez reported experimental moments for $^{13}$CO$_2$ and $^{16}$O$^{13}$C$^{18}$O \cite{Alvarez2026}. Within the  BO approximation, the polarizabilities of these isotopologues are related, providing a test for the quality of the vibrational wave functions. The transition moments were derived from the Raman spectrum of $^{13}$CO$_2$ with a small fraction of $^{16}$O$^{13}$C$^{18}$O in the range 2470--2800 cm$^{-1}$ at 298 K, as well as from the spectrum in the range 1200--1460 cm$^{-1}$ at 570 K. In addition, we have experimentally obtained the Raman spectrum of the same isotopic mixture under comparable temperature conditions. The aim of this work is to present the simulation of both spectra within the framework of the $SU_1(2) \times U(3) \times SU_2(2)$ model, together with their comparison with experiment, \textcolor{black}{explaining the origin of the differences between  the  two  experiments}.

This paper is organized as follows. Section~\ref{vibrational} presents the model used to obtain the vibrational wave functions. Section~\ref{raman} describes the procedure for calculating the Raman spectra, while in Section~\ref{experiment} a description of the experiment is summarized. Section~\ref{simulations} discusses the simulated Raman spectra as well as the associated transition moments. Finally, Section~\ref{conclusions} provides a summary and the conclusions.

\section {Vibrational description} 
\label{vibrational}

The large splitting between the fundamental bands associated with the stretching modes in the series of carbon dioxide isotopologues can be explained by the presence of Fermi interaction, as well as by the strong coupling between the local stretching oscillators of the molecule. The presence of both effects suggests a description in terms of normal coordinates, since the polyad connecting the interacting states is naturally defined within the normal mode scheme. This constitutes the traditional framework for vibrational descriptions. However, it has been shown that these systems can also be described in terms of local oscillators through a canonical transformation. The advantage of this approach lies in the possibility of introducing anharmonicities from the outset by establishing a mapping from harmonic to anharmonic potentials. To this end, we first define the stretching internal displacement coordinates $q_i = r_i - r_e,\ (i = 1,2)$, with associated momenta $p_i$. For the bending coordinates, we adopt
\begin{equation}
\label{qaqb}
S_a=r_e ~ {\bf e}_y \cdot {{\bf r}_1 \times {\bf r}_2 \over r_1 r_2}; \
\ \ \  S_b=-r_e ~ {\bf e}_x \cdot {{\bf r}_1 \times {\bf r}_2 \over r_1
r_2},
\end{equation}
where ${\bf e}_x$ and ${\bf e}_y$ are unit vectors along the $x$ and $y$ axes, respectively. In this case, coordinates carrying well-defined angular momentum are more appropriate:
\begin{equation}
\label{Qpm}
{\cal S}_\pm=\mp{1 \over \sqrt{2}} (S_a \pm i S_b); \ \ \ \ {\cal P}_\pm=\mp{1 \over
\sqrt{2}} (p_a \mp i p_b).
\end{equation}
For the stretching coordinates, we introduce the combinations
\begin{equation}
\label{Qs}
S_g={1 \over \sqrt{2}}(q_1+q_2); \quad \quad S_u={1 \over
\sqrt{2}}(q_1-q_2),
\end{equation}
which correspond to symmetry-adapted coordinates in the case of symmetric isotopologues. Equation~(\ref{Qs}) defines the transformation ${\bf S}^\dagger={\bf q}^\dagger{\bf M}$. It is then convenient to introduce new curvilinear coordinates ${\cal S}$ through the transformation ${\bf S}={\bf L}{\cal S}$, where the matrix ${\bf L}$ is determined by diagonalizing the second-order Hamiltonian according to
\begin{equation}
\label{GF}
{\bf L}^{-1} {\bf  G}_0({\bf S}) {\bf F}({\bf S}) {\bf L}={\bf \Lambda},
\end{equation}
with ${\bf  G}_0^{-1}({\bf S})={\bf M}^{\dagger}{\bf { G}}_0^{-1}({\bf q}){\bf M}$ and ${\bf F}({\bf S})={\bf M}^{\dagger}{\bf { F}}({\bf q}){\bf M}$, where ${\bf G}_0$ is the Wilson matrix evaluated at equilibrium. \textcolor{black}{ Therefore, the matrix ${\bf L}$ will be the identity matrix for the symmetric isotopologues of \ce{CO2}.}

The general form of the Hamiltonian, neglecting the potential dependence on the momenta, is given by
\begin{equation}
\label{ham}
~\hat H={1 \over 2} {\cal P}^\dagger {\bf G({\cal S}) }{\cal P} +V({\cal S}).
\end{equation}
Here, the conjugate momentum associated with the coordinate ${\cal S}_i$ is approximated as $\hat {\cal P}_i=-i \hbar \partial/\partial {\cal S}_i$ \cite{Halonen2000}. An appropriate approach for semirigid molecules consists of expanding both the Wilson matrix and the potential in power series of the coordinates ${\cal S}$, retaining only terms that conserve the polyad
\begin{equation}
\label{pN4}
P_N= 2 (\nu_1+\nu_3)+\nu_2,
\end{equation}
which accounts for the dominant Fermi resonance and Darling--Dennison-type interaction. Alternative polyads may be considered, but previous simulations indicate that this choice is suitable for describing the Raman spectra in the   energy interval of interest \cite{Bermudez2019}, a fact that will be confirmed by this work. Although a polyad-conserving Hamiltonian can be formulated in configuration space, an algebraic representation is more convenient for identifying the relevant interactions. To this end, it is important to notice that the coordinates ${\cal S}$ are curvilinear and can therefore be expressed as an expansion in terms of rectilinear normal coordinates, denoted as $\{{\cal Q}_1, {\cal Q}_3, {\cal Q}_\pm\}$. This expansion, up to cubic terms, takes the form
\begin{subequations}
	\label{curv}
	\begin{eqnarray}
		{\cal S}_1&=&{\cal Q}_1+C_1\, {\cal Q}_1{\cal Q}_+{\cal Q}_-+C_2\, {\cal Q}_3{Q}_+{\cal Q}_-+C_3\, {\cal Q}_+{\cal Q}_-; \\[0.15 cm]
		{\cal S}_{+}&=&{\cal Q}_{+}+ C_4\, {\cal Q}_{+}{\cal Q}_1+C_5\, {\cal Q}_{+}{\cal Q}_3+C_6\, {\cal Q}_{+}{\cal Q}_1^2 \nonumber \\
		&+&C_7\, {\cal Q}_{+}{\cal Q}_3^2+C_8\, {\cal Q}_{+}{\cal Q}_1{\cal Q}_3-2\, C_9\,{\cal Q}_{+}^2{\cal Q}_{-}; \\[0.15 cm]
		{\cal S}_{-}&=&{\cal Q}_{-}+ C_4\, {\cal Q}_{-}{\cal Q}_1+C_5\, {\cal Q}_{-}{\cal Q}_3+C_6\, {\cal Q}_{-}{\cal Q}_1^2 \nonumber \\
		&+&C_7\, {\cal Q}_{-}{\cal Q}_3^2+C_8\, {\cal Q}_{-}{\cal Q}_1{\cal Q}_3-2\, C_9\, {\cal Q}_{+}{\cal Q}_{-}^2; \\[0.15 cm]
		{\cal S}_3&=&{\cal Q}_3+C_{10}\, {\cal Q}_3{\cal Q}_+{\cal Q}_-+C_{11}\, {\cal Q}_1{\cal Q}_+{\cal Q}_-+C_{12}\, {\cal Q}_+{\cal Q}_-,
	\end{eqnarray}
\end{subequations}
where the coefficients $\{C_i;\ i=1,\dots, 12\}$ depend on the molecular structure and force constants, as well as on the transformation matrix ${\bf L}$ \cite{suarez2026}. In the limit of small oscillations, the curvilinear coordinates reduce to the rectilinear normal coordinates, as expected.

The algebraic realization of the coordinates (\ref{curv}) is obtained by introducing a bosonic representation for the stretching normal modes,
\begin{equation}
\label{Aiota}
{\cal A}_\varsigma^\dagger={1 \over \sqrt{2}} \left( \beta_\varsigma {\cal 
Q}_\varsigma-{i \over \hbar \beta_\varsigma} {\cal P}_\varsigma\right); \quad {\cal 
A}_\varsigma={1 \over \sqrt{2}} \left( \beta_\varsigma {\cal Q}_\varsigma+{i \over \hbar 
\beta_\varsigma} {\cal P}_\varsigma\right),
\end{equation}
where
$\beta^2_\varsigma={1 \over \hbar} (\lambda_\varsigma)^{1/2}$ with 
$\lambda_\varsigma={ \Lambda}_{\varsigma \varsigma}$, while for the bending coordinates
\begin{equation}
\label{Atau}
{\tau}^\dagger_\pm={1 \over \sqrt{2}} \left( \sqrt{\mu \omega  \over \hbar}{ 
{Q}}_\pm + i {  P_\mp \over \sqrt{\hbar \mu \omega}}\right); \quad {\tau}_\mp=-{1 \over \sqrt{2}} \left( \sqrt{\mu \omega  \over \hbar}{ 
{Q}}_\pm- i {  P_\mp \over \sqrt{\hbar \mu \omega}}\right),
\end{equation}
with $\omega=\sqrt{g^{(0)}_{aa} f_{aa}}$ and $\mu=1/g^{(0)}_{aa}$.
Taking these transformations into account, an algebraic realization of the curvilinear coordinates is obtained. In principle, the Hamiltonian is expanded in terms of curvilinear coordinates, and its algebraic expression follows from the substitution of (\ref{curv}), together with (\ref{Aiota}) and (\ref{Atau}). This procedure can be simplified by retaining only the linear terms in the expansions (\ref{curv}). Such an approximation is justified within the framework of an effective Hamiltonian, where the spectroscopic parameters effectively renormalize higher-order contributions. Following this approach, we have found that the Hamiltonian capable of providing spectroscopic-quality descriptions for the complete set of isotopologues is given by \cite{marisol22,marisol24}
\begin{eqnarray}
\label{H1locLN3}
\hat H&=& \sum_{\varsigma=1,3}
   {\hbar  \Omega_\varsigma \over 2} ({\cal A}_\varsigma^\dagger {\cal 
A}_\varsigma+{\cal A}_\varsigma {\cal A}_\varsigma^\dagger)+ {\hbar \omega \over 2} 
\sum_{\sigma=+,-} (\tau^\dagger_\sigma \tau_\sigma+\tau_\sigma 
\tau^\dagger_\sigma)+\sum_{i,j} x_{ij} \{\hat \nu_i, \hat \nu_j \}\nonumber \\
&+& 
g_{22} {\ell}^2+\alpha_F [{\cal A}_1^\dagger  \tau_+ \tau_-+{\rm H.c.} \big]+\alpha^{(1)}_F \{  \hat \nu_1, \hat V_F \}+\alpha^{(3)}_F 
\{  \hat \nu_3, \hat V_F \}+d_N \hat D,\nonumber \\
\end{eqnarray}
where $
\Omega_\varsigma=\lambda_\varsigma^{1/2}$ and  $\{  \hat A, \hat B\}={1 \over 2}( \hat A \hat B+ \hat B \hat A )$. 
In this expression, $\hat \nu_\varsigma$ denotes the number operator associated with the $\varsigma$-th normal mode. The Fermi and Darling-Dennison interactions take the form
\begin{equation}
\label{hN2}
\hat V_F=\alpha_F  [{\cal A}^\dagger_1 \tau_+ 
\tau_- +{\rm H.c.}]; \ \ \ \ \ \ \hat D={\cal A}^\dagger_1 {\cal A}^\dagger_1 {\cal A}_3 {\cal A}_3+{\rm H.c.}.
\end{equation}
Up to this point, we have followed the conventional effective Hamiltonian formulation.

To improve the description, we introduce for the stretching modes the canonical transformation \cite{suarez2026,Suarezn2o}
\begin{subequations}
\label{CT}
\begin{eqnarray}
{\cal A}_1^{\dagger}&=& \alpha\, c^\dagger_1+\sqrt{1-\alpha^2}\, c^\dagger_2; \\
{\cal A}_3^{\dagger}&=& \sqrt{1-\alpha^2}\, c^\dagger_1-\alpha\,  
c^\dagger_2, 
\end{eqnarray}
\end{subequations}
where the operators $c_i^\dagger(c_i)$ are bosonic operators isomorphic to the local bosonic operators $a_i^\dagger(a_i)$. The coefficient $\alpha$ depends on the force and structural constants as demonstrated in Ref.~\cite{Suarezn2o}. The incorporation of this transformation into the Hamiltonian preserves the polyad and yields the same eigensystem. The crucial step in our approach consists in establishing the mapping
\begin{equation}
\label{anar}
c_i^\dagger(c_i) \to b_i^\dagger(b_i),
\end{equation}
where the new operators are $SU(2)$ ladder operators $b_i^\dagger(b_i)$ associated with the Morse states $|{j} {v_i}\rangle$:
\begin{equation}\nonumber
b^\dagger_i |j_iv_i \rangle=k_+
|j_iv_i+1 \rangle; \ \ \ \ 
b_i|j_iv_i\rangle=k_-
|j_iv_i-1 \rangle,
\end{equation}
with $k_+=\sqrt{(v_i+1)[1-(v_i+1)/\kappa_i]}$ and $k_-=\sqrt{v_i(1-v_i/\kappa_i)}$. Here $v_i=0,1,\dots j_i-1$ denotes the vibrational quantum number, and $\kappa_i=2
j_i+1$ is related to the depth of the potential. These parameters are connected for different isotopologues through the relation
\begin{equation}
\label{relk}
\kappa_i=\sqrt{{g^{(0)}_{jj} \over g^{(0)}_{ii} }} \kappa_j; \ \ \ i,j=1,2.
\end{equation}
where $g^{(0)}_{jj}$ is the element of the Wilson matrix evaluated at equilibrium for the $j$-th stretching local oscillator. The parameters $\kappa_i$ for the whole set of isotopologues are obtained by taking the principal isotopologue as a reference.

Because of the degenerate nature of the bending modes, the anharmonization for these degrees of freedom is carried out within the framework of the $U(3)$ model. The model consists of truncating the space of two degenerate bosonic operators $\tau^\dagger_\pm(\tau_\pm)$ through the introduction of a scalar bosonic operator $s (s^\dagger)$, together with the constraint of a fixed total number of bosons $\hat N=\hat n+\hat n_s$. In this way, the total number of bosons $N$ determines the dimension of the Hilbert space associated with the bending degrees of freedom. In this case, the anharmonization is implemented through the mapping $\tau_\pm^\dagger(\tau_\pm) \to b^\dagger_\pm (b_\pm)$, with the following action on the 2D harmonic oscillator states
\begin{equation}
\label{ladd-opU3}
b^\dagger_\pm|[N];n^{\ell}\rangle=\eta_+ \
|[N];(n+1)^{{\ell}\pm 1} \rangle; \ \ \  
b_\pm |[N];n^{\ell}  \rangle=\eta_-|[N];(n-1)^{{\ell} \mp 1}\rangle,
\end{equation}
where $\eta_+=\sqrt{\left(\frac{n\pm{\ell}}{2}+1\right)\left(1-\frac{n}{N}\right)}$ and $\eta_-=\sqrt{\left( \frac{n \pm {\ell}}{2} 
\right) \left(1-\frac{n-1}{N} \right)}$, with $\ell$ denoting the angular momentum projection. In a similar manner to the relation (\ref{relk}), it is possible to derive the following expression \cite{marisol24} 
\begin{equation}
\label{relNR}
N_i=\sqrt{{1/16+1/16+4/12 \over 1/m_{O_1}+1/m_{O_2}+4/m_C }} N,
\end{equation}
where $N$ corresponds to the value obtained for the principal isotopologue. The model presented here is known as the $SU_1(2) \times U(3) \times SU_2(2)$ model, since this product of groups defines the dynamical group of a triatomic linear molecule.  

The anharmonization procedure, once the transformation (\ref{CT}) and the operators of Eq.~(\ref{ladd-opU3}) is incorporated into the Hamiltonian, provides a more efficient model for describing vibrational excitations, leading to improved wave functions compared to the traditional harmonic oscillator basis. The vibrational states are constructed starting from the local basis
\begin{equation}
\label{LocL}
    |P;j_1,j_2,[N];v_1 v_2 n^\ell\rangle=|j_1v_1 \rangle \otimes |j_2 v_2 \rangle \otimes |[N];n_+ n_- \rangle
\end{equation}
where $|[N];n_+ n_- \rangle$ corresponds to the states associated with the $U(3) \supset U(2)$ basis, with $n_+= (n + \ell)/2$ and $n_-= (n - \ell)/2$. The local basis is organized into sets with a well-defined polyad number (\ref{pN4}), which explains the label $P$ in Eq.~(\ref{LocL}). This basis is projected onto states carrying irreducible representations of the group ${\cal D}_{\infty h}$ with labels $\{ \Gamma,\gamma=\pm\ell \}$, according to the group chain ${\cal D}_{\infty h} \supset {\cal C}_{\infty v}$: 
\begin{equation}
\label{base}
|q,\Gamma(\pm\ell),P \rangle=\sum_{v_1,v_2, n_+,n_-} C^{(q,\Gamma,P)}_{v_1,v_2, n_+,n_-}|P;j_1,j_2,[N];v_1 v_2 n_+ n_-\rangle.
\end{equation}  
 In this expression, the label $q$ denotes a multiplicity index. To obtain a basis isomorphic to the normal-mode representation, the number operators $\hat \nu_\varsigma={\cal A}^\dagger_\varsigma {\cal A}_\varsigma$ and $\hat n=\tau^\dagger_+ \tau_+ + \tau^\dagger_- \tau_- $ are diagonalized in the harmonic basis, yielding states with well-defined normal quantum numbers $\{ \nu_1,\nu_2,\nu_3\}\equiv\{ \nu \}$. 
\begin{equation}
\label{basenOR}
|\{\nu \};\Gamma(\pm\ell),P \rangle=\sum_q A_{q,{\nu}}|q,\Gamma(\pm\ell),P \rangle.
\end{equation} 
These states can, in turn, be expressed in the original local basis
\begin{equation}
\label{baseloc}
|\{\nu \};\Gamma(\pm\ell),P \rangle=\sum_{v_1,v_2, n_+,n_-} D^{(\nu,\Gamma,P)}_{v_1,v_2, n_+,n_-}|P;j_1,j_2,[N];v_1 v_2 n_+ n_-\rangle,
\end{equation}
with $D^{(\nu,\Gamma,P)}_{v_1,v_2, n_+,n_-}=
\sum_q A_{q,{\nu}} C^{(q,\Gamma,P)}_{v_1,v_2, n_+,n_-}$. The diagonalization of the Hamiltonian leads to eigenstates of the form
\begin{equation}
\label{basenu}
|\beta;\Gamma(\pm\ell),P \rangle=\sum_{\{\nu \}} B_{\{ \nu \},\beta}\, |\{\nu \};\Gamma(\pm\ell),P \rangle,
\end{equation}
which provide the vibrational states used to describe the Raman spectra. Here, the index $\beta$ labels the energy $E_{\beta}$.

\section {Raman spectroscopy}
\label{raman}

Raman spectroscopy is a two-photon inelastic scattering process in which a photon of energy $h \nu_0$ is absorbed and a photon of different energy $h \nu'$ is emitted, the latter being related to the energy difference $h \Delta\nu$ involved in the molecular transition by $h \nu' = h \nu_0 - h \Delta\nu$. The isotropic contribution of the polarizability to the Raman intensities is obtained from the differential cross section [$\mathrm{m^2/sr}$] for the Q-branches ($\Delta J=0$), which is given for a gas sample at thermal equilibrium~\cite{Alvarez2024} as
\begin{equation}
    \label{RTI}
    \bigg( {d \sigma \over d\Omega} \bigg)^{\rm trace}_{i \to f}=\bigg({\pi \over \epsilon_0} \bigg)^2 {(\bar\nu_0-\Delta\bar\nu)^4 \over Z(T)}\,(2J+1)\,g_{\rm ns}\, g_\nu\, |M_{if}|^2 \exp\bigg({-hc\, E_{\nu,J} \over k_B T}\bigg), 
\end{equation}
 where $\epsilon_0$ is the vacuum permittivity, $\bar\nu_0\ \mathrm{[cm^{-1}]}$ is the wavenumber of the exciting radiation, the Raman shift is $\Delta\bar\nu=h\Delta\nu=E_{\nu',J}-E_{\nu,J}$, with $E_{\nu,J}$ and $E_{\nu',J}$ denoting the ro-vibrational term values of the initial and final states, respectively (in $\mathrm{cm^{-1}}$), which in this work are approximated as $E_{\nu,J}=G_\nu + B_\nu\, [J(J+1)-\ell^2]$. The factors $g_{\mathrm{ns}}$ and $g_\nu$ stand for the nuclear spin and vibrational degeneracies of the initial state, and $Z(T)$ is the internal partition function at temperature $T\ \mathrm{[K]}$. In this work, $Z(T)$ is taken from HITRAN2024, where it is computed using the \textcolor{black}{product approximation, in which the internal partition function is factorized into its vibrational and rotational contributions \cite{Gamache2025}.} The intensities are determined by the transition moments $M_{if}=\langle \nu_i|\bar \alpha| \nu_f \rangle$ $\mathrm{[C\cdot m^2/V]}$ of the totally symmetric mean polarizability $\bar \alpha$. In this work, we confine the simulations to  the isotropic contribution. {Anisotropic contributions are expected to be of lower order, especially for the Q-branches of the spectrum, as discussed in Ref.~\cite{Alvarez2024}. \textcolor{black}{It is important to notice that, within the rigid-rotor approximation, the Raman shift is given by $\Delta\bar\nu = G_{\nu'} - G_\nu$ and does not depend on the rotational quantum number.}

In order to calculate the transition moments, the polarizability is expanded around the molecular equilibrium configuration in terms of the curvilinear coordinates ${\cal S}$, which, up to cubic terms, takes the form  
\begin{eqnarray}
	\label{alpha}
    \bar\alpha &=& \bar \alpha_0+\sum_{\varsigma=1,3} \bigg( {\partial \bar \alpha \over \partial{\cal S}_\varsigma}\bigg)_0 {\cal S}_\varsigma +{1 \over 2}\sum_{\varsigma=1,3} \bigg( {\partial^2 \bar \alpha \over \partial {\cal S}_\varsigma^2} \bigg)_0 {\cal S}_\varsigma^2 \nonumber \\
    &+&\bigg({\partial^2 \bar \alpha \over \partial {\cal S}_1 \partial {\cal S}_3} \bigg)_0 {\cal S}_1 {\cal S}_3+\bigg( {\partial^2 \bar \alpha \over \partial{\cal S}_+\partial {\cal S}_-}  \bigg)_0 {\cal S}_+{\cal S}_-\nonumber\\
    &+&\sum_{\varsigma=1,3} \bigg({ \partial^3 \bar \alpha \over  \partial {\cal S}_\varsigma \partial{\cal S}_+\partial {\cal S}_-}  \bigg)_0 {\cal S}_\varsigma{\cal S}_+{\cal S}_-.
\end{eqnarray}
For symmetric isotopologues,  the stretching derivatives in Eq.~(\ref{alpha}) simplify to $\partial/\partial {\cal S}_1=\partial/\partial S_g$ and $\partial/\partial {\cal S}_3=\partial/\partial S_u$. It should be stressed that, in the symmetric limit, not all the terms involved in (\ref{alpha}) survive due to symmetry considerations. Hence, Raman spectra for asymmetric isotopologues are expected to exhibit a more complex structure.

The expansion (\ref{alpha}) is not suitable for the calculation of the matrix elements using the current approach, since the wave functions are defined within the framework of the $SU_1(2) \times U(3) \times SU_2(2)$ algebraic model. The appropriate route consists of substituting the curvilinear coordinates in terms of rectilinear normal coordinates given by (\ref{curv}). Once this substitution is performed, an algebraic realization compatible with the model is obtained through (\ref{Aiota}) for the stretching coordinates and (\ref{Atau}) for the bending coordinates, followed by the application of the canonical transformation (\ref{CT}) and the anharmonization procedure (\ref{anar}) and (\ref{ladd-opU3}). Following this methodology, the mean polarizability is expressed in terms of the generators of the dynamical algebra together with the potential parameters:
\begin{equation}
	\hat{\alpha}_{\Sigma^+}=\bar{\alpha}(\kappa_i,[N];\hat{b}_j^\dagger,\hat{b}_j,\hat{b}_\pm^\dagger,\hat{b}_\pm)\ ; \qquad j=1,2.
\end{equation}

To compute the transition moments $M_{if}$, both the derivatives of the polarizability and the vibrational wave functions are required. We have estimated the derivatives  by fitting $42$ experimental transition moments reported for the main isotopologue of CO$_2$ \cite{Alvarez2024,tejeda,chrysos} in the range 1200--4700 cm$^{-1}$, using the wave functions obtained from the vibrational description given in Refs.~\cite{marisol17,marisol22,marisol24}. The fitting procedure consists of minimizing a root-mean-square deviation, where the derivatives are treated as fitting parameters, as described in detail in Ref.~\cite{Bermudez2019,suarez2026}. The values of the derivatives depend strongly on the connection between normal and local modes. In this work, the canonical transformations (\ref{CT}) are employed, whereas in Ref.~\cite{suarez2026} a Bogoliubov-type transformation was used. The fitted values of the derivatives are presented in Table~\ref{deriv}, while the comparison between computed and experimental transition moments can be found in the supplementary material of Ref.~\cite{suarez2026}.
\begin{table}[h!]
    \centering
    \scriptsize
    \caption{Mean polarizability derivatives with respect to symmetry-adapted curvilinear coordinates (\ref{Qpm}) and (\ref{Qs}). A set of 42 experimental transition moments from Refs.~\cite{Alvarez2024,tejeda,chrysos} for the most abundant isotopologue was fitted following the method described in Ref.~\cite{suarez2026}.}

    \
    
    \begin{tabular}{ccccc}
        \toprule
        $\displaystyle\left(\frac{\partial\bar{\alpha}}{\partial S_g}\right)_0$ & $\displaystyle\left(\frac{\partial^2\bar{\alpha}}{\partial S_g^2}\right)_0$ & $\displaystyle\left(\frac{\partial^2\bar{\alpha}}{\partial S_u^2}\right)_0$ & $\displaystyle\left(\frac{\partial^2\bar{\alpha}}{\partial {\cal S}_+\partial {\cal S}_-}\right)_0$ & $\displaystyle\left(\frac{\partial^3\bar{\alpha}}{\partial S_g\partial {\cal S}_+\partial {\cal S}_-}\right)_0$ \\ [0.3 cm]
        ($10^{-30}$ CV$^{-1}$m) & ($10^{-20}$ CV$^{-1}$) & ($10^{-20}$ CV$^{-1}$) & ($10^{-20}$ CV$^{-1}$) & ($10^{-10}$ CV$^{-1}$m$^{-1}$) \\
         \midrule
        3.22 & 2.56 & 0.47 & 0.23 & 1.77 \\
         \bottomrule
    \end{tabular}
    \label{deriv}
\end{table}

Assuming the BO approximation, the fitted derivatives can be used to extrapolate the corresponding derivatives with respect to the mass dependent coordinates for any isotopologue. It should be noted that, while for the symmetric isotopologues the polarizability derivatives are mass independent, for asymmetric isotopologues a mass dependence arises, as can be seen from the chain rule
\begin{equation}
	\label{chain}
	\frac{\partial}{\partial {\cal S_\varsigma}}=\sum_{\alpha=1}^2 { L}_{\alpha\varsigma}\frac{\partial}{\partial S_\alpha}, 
\end{equation}
for the $ \varsigma$-th normal mode and  $\alpha=g,u$. This relation allows us to express mass-dependent derivatives in terms of mass-independent coordinates for any isotopologue. In Table~3 of Ref. \cite{suarez2026} the extrapolated polarizability surfaces for the series of isotopologues are displayed. The derivatives, together with the wave functions, provide the necessary ingredients to carry out the simulation of the Raman spectra. We next present the experimental set up for the Raman spectra of the the isotopologues we shall simulate. 

\section {Experimental acquisition of spectra of CO$_2$ at 500 and 650~K} 
\label{experiment}

For the experimental Raman analyses of carbon dioxide in the gas phase, the commercially available standard gases natural carbon dioxide with 98.9~\% \ce{^12C} (Air Liquide) or an \ce{^13C} enriched CO$_2$ (99.0~\% \ce{^13C}, Sigma-Aldrich 364592) were used. The gas was introduced into an evacuated small fused silica capillary with an outer diameter of 320~$\mu$m and an inner diameter of 75~$\mu$m by a high-pressure syringe pump to a final pressure of 1~MPa. The capillary was introduced from the side into a Linkam CAP500 heating chamber, where the capillary could be temperature-controlled up to 650~K with an accuracy of better than 2~K in a silver heating block. Through a slit opening in the heating block the capillary optically was accessible to the laser and the back-scattered Raman signal could be collected. The analyses were performed at different temperatures, here the data for 500 and 650~K are shown.

The Raman spectra were acquired as unpolarized spectra in backscattering geometry using a Horiba Jobin Yvon LabRam HR800 UV-Vis Evolution system. A green VERDI V2 laser with a wavelength of 532.131~nm was used as energy source with an output power of 2.020~W, of which ca. 1.2~W reached the top of the fused silica capillary. The Raman signal was collected with a 50x Nikon objective through a very small pinhole of 25~µm, and on passing the 800~mm focal length dispersed by a 2400 grooves/mm grating on a back-illuminated Symphony CCD with 2048$\times$512 pixels cooled by liquid nitrogen to $-$133~°C. This resulted in a pixel resolution on the CCD of less than 0.046~nm or less than 0.14~cm$^{-1}$. Each spectrum was measured for 300~s with three accumulations.

\section {Spectra simulations} 
\label{simulations}

Since the eighties, several approaches to simulate Raman spectra of the main isotopologue of carbon dioxide have been reported, albeit limited to energies below $4000$ cm$^{-1}$ \cite{Papineau1983}. In a later work, Sepman {\it et al.} presented fits to measured Raman spectra for several diatomic and triatomic molecules, including CO$_2$. The Raman spectrum of CO$_2$ was simulated at $T=1790$ K, restricted to the range 1320--1520 cm$^{-1}$ \cite{Sepman2013}. Subsequently, using the algebraic approach described in this contribution, restricted to pure vibrational degrees of freedom, our group reported the simulation of the Raman spectrum of \ce{^12C^16O2} at $T=1743$ K in the range 1150--1500 cm$^{-1}$, first limited to the polyad $P_{212}=2(\nu_1+\nu_3)+\nu_2$ \cite{Lemus2014}, and later extended to the polyads $P_{213}=2 \nu_1+\nu_2+3 \nu_3$ and $P_{214}=2 \nu_1+\nu_2+4 \nu_3$ \cite{Bermudez2019}. In subsequent work, Lill \textit{et al.} described the isotopologues \ce{^12C^16O2}, \ce{^13C^16O2}, \ce{^16O^12C^17O}, and \ce{^16O^12C^18O} in the interval $296$--$2355$ K, taking into account both isotropic and anisotropic scattering,  including an extensive set of ro-vibrational transitions \cite{Lill1}.

In a recent work, \'Alvarez {\it et al.} obtained the polarizability transition moments of \ce{^12C^16O2} by measuring Raman intensities at known temperatures, the latter being determined from the rotational spectrum~\cite{Alvarez2024}. Their analysis provided a total of 38 transitions, from which 34 correspond to new polarizability transition moments of hot bands involving energies up to 5000~cm$^{-1}$. Using our approach, in this work we have computed the transition moments employing the wave functions (\ref{basenu}) associated with the polyad $P_{212}$ and provided by the fits reported in Refs.~\cite{marisol17,marisol22,marisol24}. These transition moments are consistent with the experimental ones and enable the simulation of Raman spectra at arbitrary temperatures.

As a test of our model, we simulate \textcolor{black}{in the rigid-rotor approximation} the Raman spectrum of \ce{^12C^16O2} at 1780~K and compare it with the experimental spectrum reported in Fig.~3 of Ref.~\cite{Alvarez2024}. Previously, the same spectrum was simulated using a limited set of transition moments taken from Ref.~\cite{Lemus2014}, which results in the absence of several hot bands. In contrast, the present simulation includes an extended set of 1692 transitions between states with energies up to $21\,400$~cm$^{-1}$, allowing for a more complete description of the hot-band structure, \textcolor{black}{even at higher temperatures}. The resulting spectrum is shown in Fig.~\ref{1780K}, where it can be observed that our simulation shows good agreement in the interval 1200--1300~cm$^{-1}$; however, significant discrepancies in the intensities arise in the higher-wavenumber region, particularly in the range 1360--1500~cm$^{-1}$. Despite these discrepancies, all hot bands are reproduced in the simulated spectrum. The differences between the experimental and simulated spectrum  arises from the rotational structure, as will be discussed later.

\begin{figure}[h!]
    \centering
\includegraphics[width=0.8\textwidth]
{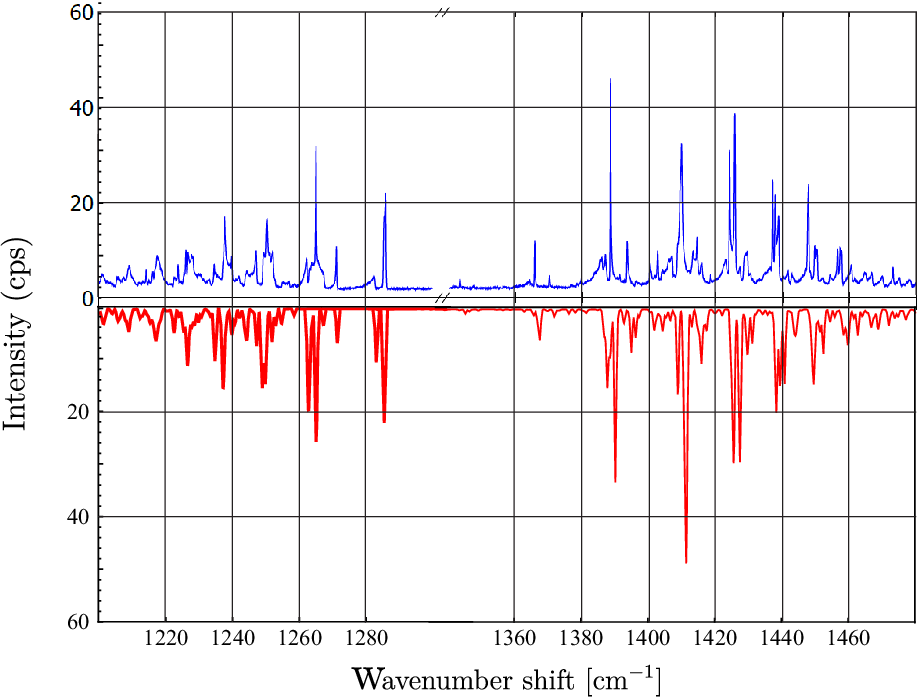}
    \caption{Experimental Raman spectra (top panel, in blue) compared with the pure vibrational simulation (bottom panel, in red) for the most abundant isotopologue at $T = 1780$~K. The experimental data were taken from Ref.~\cite{Alvarez2024}. Wavenumber shifts were calculated using vibrational energies from Ref.~\cite{marisol17}, assuming Gaussian peak profiles with a full width at half maximum of 0.7~cm$^{-1}$.}
    \label{1780K}
\end{figure}

More recently, \'Alvarez {\it et al.} \cite{Alvarez2026} obtained a set of experimental mean polarizability transition moments for the isotopologues $^{13}$C$^{16}$O$_2$ and $^{16}$O$^{13}$C$^{18}$O using the same approach previously followed for the principal isotopologue. In principle, assuming the BO approximation, if the derivatives of the polarizability with respect to isotopically invariant coordinates are known, the transition moments of different isotopologues are connected as long as their vibrational wave functions are available. This relation is indeed possible within our approach, since the vibrational description has been carried out for nine isotopologues. It is important to note that, in practice, this relation is not exact due to the inherent approximations involved in obtaining the wave functions, which implies that any model can be tested through the connection of the transition moments. In this context, it is possible to obtain  the polarizability derivatives by taking into account experimental transition moments for different isotopologues. This approach may provide improved  derivatives that are not well determined from the transitions available for a single isotopologue. This analysis is currently in progress.

The estimation of the transition moments in Ref.~\cite{Alvarez2026} was performed by considering the Raman intensities of a mixture of the $^{13}$C$^{16}$O$_2$ (636) and $^{16}$O$^{13}$C$^{18}$O (638) isotopologues in the region $1200$--$1460$ cm$^{-1}$, associated with the Fermi resonance, at $570$ K. An additional Raman spectrum for $^{13}$C$^{16}$O$_2$ was obtained in the region $2470-2800$ cm$^{-1}$ at $298.3$ K. Using the polarizability derivatives previously fitted for the principal isotopologue we obtain the transition moments  displayed in Tables~\ref{tm636} and \ref{tm638} together with the comparison to the experimental results. In these tables, the transitions are labeled according to the dominant component $B_{{\nu},\beta}$ of the wave functions in the normal basis set, as given in Eq.~(\ref{basenu}).  Most of the residuals are smaller than the experimental uncertainty, indicating that the transition moments are well determined. The theoretical transition moments are thus quite reasonable, confirming the good quality of both the wave functions and the polarizability surface.

{\scriptsize
\begin{longtable}{P{1.1cm}P{2.1cm}P{1.5cm}*4{P{0.75cm}}}
	\caption{Comparison between calculated $|M^{(\mathrm{calc})}|$ and experimental transition moments $|M^{(\mathrm{exp})}|$ (in $10^{-42}\ \mathrm{CV^{-1}m^2}$) taken from Ref.~\cite{Alvarez2026} for the isotopologue $^{13}$C$^{16}$O$_2$. Here, $\Delta\bar\nu$ denotes the wavenumber shift of the scattered radiation, obtained from vibrational energies reported in Ref.~\cite{marisol22}, $\sigma$ the experimentally determined uncertainty \cite{Alvarez2026}, and $\Delta$ the residual in the transition moment with respect to the experimental values, while the rms deviation is calculated using the definition given in Eq.~(22) of Ref.~\cite{suarez2026}.} \label{TMET} \\
	
	\toprule
	Band \# & $|\nu_i\rangle\to |\nu_f\rangle$ & $\Delta\bar\nu$ [cm\textsuperscript{$-1$}] & $|M^{(\mathrm{exp})}|$ & $\sigma$ & $|M^{(\mathrm{calc})}|$ & $\Delta$  \\ \midrule
	\endfirsthead	
	\multicolumn{7}{l}%
	{{\tablename\ \thetable{}} (continued)} \\
	\toprule
	Band \# & $|\nu_i\rangle\to |\nu_f\rangle$ & $\Delta\bar\nu$ [cm\textsuperscript{$-1$}] & $|M^{(\mathrm{exp})}|$ & $\sigma$ & $|M^{(\mathrm{calc})}|$ & $\Delta$  \\ \midrule 
	\endhead
	\bottomrule \multicolumn{7}{r}{{(continued on next page)}} \\ 
	\endfoot
	\bottomrule
	\endlastfoot

    1 & $|04^20\rangle\to|14^20\rangle$ & $1220.63$ & $5.63$ & $0.92$ & $6.04$ & $-0.41$ \\
    2 & $|03^30\rangle\to|05^30\rangle$ & $1222.97$ & $4.13$ & $0.26$ & $4.46$ & $-0.33$ \\
    3 & $|03^10\rangle\to|13^10\rangle$ & $1230.72$ & $5.75$ & $0.39$ & $5.94$ & $-0.19$ \\
    4 & $|02^20\rangle\to|04^20\rangle$ & $1234.35$ & $4.48$ & $0.21$ & $4.49$ & $-0.01$ \\
    5 & $|12^20\rangle\to|22^20\rangle$ & $1236.92$ & $7.03$ & $1.00$ & $5.17$ & $\ms1.86$ \\
    6 & $|02^00\rangle\to|04^00\rangle$ & $1241.66$ & $5.60$ & $0.20$ & $5.62$ & $-0.02$ \\
    7 & $|01^10\rangle\to|03^10\rangle$ & $1248.03$ & $4.50$ & $0.19$ & $4.47$ & $\ms0.03$ \\
    8 & $|11^10\rangle\to|21^10\rangle$ & $1252.65$ & $5.01$ & $0.26$ & $5.43$ & $-0.42$ \\
    9 & $|00^00\rangle\to|02^00\rangle$ & $1265.86$ & $4.32$ & $0.14$ & $4.29$ & $\ms0.03$ \\
    10\textsuperscript{*} & $|10^00\rangle\to|20^00\rangle$ & $1275.03$ & $6.05$ & $0.69$ & $5.97$ & $\ms0.08$ \\
    11 & $|00^01\rangle\to|10^01\rangle$ & $1349.49$ & $7.22$ & $0.40$ & $7.95$ & $-0.73$ \\
    12 & $|00^00\rangle\to|10^00\rangle$ & $1369.98$ & $7.71$ & $0.16$ & $7.88$ & $-0.17$ \\
    13 & $|20^00\rangle\to|30^00\rangle$ & $1375.02$ & $7.00$ & $3.00$ & $11.19$ & $-4.19$ \\
    14 & $|02^00\rangle\to|20^00\rangle$ & $1379.14$ & $7.99$ & $0.24$ & $8.09$ & $-0.10$ \\
    15\textsuperscript{*} & $|10^00\rangle\to|12^00\rangle$ & $1380.46$ & $9.83$ & $0.58$ & $10.32$ & $-0.49$ \\
    16 & $|01^10\rangle\to|11^10\rangle$ & $1388.48$ & $7.86$ & $0.33$ & $7.78$ & $\ms0.08$ \\
    17 & $|04^00\rangle\to|22^00\rangle$ & $1391.23$ & $7.33$ & $0.42$ & $8.11$ & $-0.78$ \\
    18 & $|03^10\rangle\to|21^10\rangle$ & $1393.10$ & $7.52$ & $0.23$ & $8.02$ & $-0.50$ \\
    19 & $|12^00\rangle\to|14^00\rangle$ & $1395.28$ & $8.01$ & $0.54$ & $11.96$ & $-3.95$ \\
    20 & $|11^10\rangle\to|13^10\rangle$ & $1396.57$ & $10.62$ & $0.32$ & $10.54$ & $\ms0.08$ \\
    21 & $|02^20\rangle\to|12^20\rangle$ & $1402.84$ & $7.91$ & $0.50$ & $7.77$ & $\ms0.14$ \\
    22 & $|04^20\rangle\to|22^20\rangle$ & $1405.41$ & $6.86$ & $0.38$ & $7.97$ & $-1.11$ \\
    23 & $|12^20\rangle\to|14^20\rangle$ & $1409.33$ & $9.00$ & $3.00$ & $10.66$ & $-1.67$ \\
    24 & $|03^30\rangle\to|13^30\rangle$ & $1414.88$ & $7.66$ & $0.23$ & $7.78$ & $-0.12$ \\
    25 & $|04^40\rangle\to|14^40\rangle$ & $1425.39$ & $7.29$ & $0.34$ & $7.80$ & $-0.52$ \\
    26 & $|01^10\rangle\to|13^10\rangle$ & $2478.75$ & $0.09$ & $0.04$ & $0.08$ & $\ms0.01$ \\
    27 & $|00^00\rangle\to|04^00\rangle$ & $2507.52$ & $0.070$ & $0.020$ & $0.061$ & $\ms0.009$ \\
    28 & $|01^10\rangle\to|21^10\rangle$ & $2641.13$ & $0.10$ & $0.03$ & $0.12$ & $-0.02$ \\
    29 & $|00^00\rangle\to|20^00\rangle$ & $2645.01$ & $0.123$ & $0.018$ & $0.123$ & $\ms0.000$ \\
    30 & $|00^00\rangle\to|12^00\rangle$ & $2750.44$ & $0.062$ & $0.015$ & $0.053$ & $\ms0.009$ \\
    \midrule
    rms &  &  &  &  & $0.130$ &

    \label{tm636}
\end{longtable}}

\

\begin{table}[h!]
    \centering
    \caption{\scriptsize Comparison between calculated $|M^{(\mathrm{calc})}|$ and experimental transition moments $|M^{(\mathrm{exp})}|$ (in $10^{-42}\ \mathrm{CV^{-1}m^2}$) taken from Ref.~\cite{Alvarez2026} for the isotopologue $^{16}$O$^{13}$C$^{18}$O. Here, $\Delta\bar\nu$ denotes the wavenumber shift of the scattered radiation, obtained from vibrational energies reported in Ref.~\cite{marisol24}, $\sigma$ the experimentally determined uncertainty \cite{Alvarez2026}, and $\Delta$ the residual in the transition moment with respect to the experimental values, while the rms deviation is calculated using the definition given in Eq.~(22) of Ref.~\cite{suarez2026}.}
    \label{tm638}
    
    \
    
    \scriptsize
    \begin{tabular}{ccccccc}
        \hline
        Band \# & $|\nu_i\rangle\to |\nu_f\rangle$ & $\Delta\bar\nu$ [cm\textsuperscript{$-1$}] & $|M^{(\mathrm{exp})}|$ & $\sigma$ & $|M^{(\mathrm{calc})}|$ & $\Delta$ \\
        \hline
        a & $|02^20\rangle\to|04^20\rangle$ & $1211.54$ & $5.55$ & $0.60$ & $4.68$ & $\ms0.87$ \\
        b & $|01^10\rangle\to|03^10\rangle$ & $1225.85$ & $5.24$ & $1.18$ & $4.85$ & $\ms0.39$ \\
        c & $|00^00\rangle\to|02^00\rangle$ & $1294.92$ & $5.23$ & $0.18$ & $5.02$ & $\ms0.21$ \\
        d & $|00^00\rangle\to|10^00\rangle$ & $1342.34$ & $7.01$ & $0.54$ & $7.22$ & $-0.21$ \\
        e & $|10^00\rangle\to|12^00\rangle$ & $1359.60$ & $9.47$ & $1.05$ & $9.43$ & $\ms0.04$ \\
        f & $|01^10\rangle\to|11^10\rangle$ & $1362.12$ & $8.04$ & $1.36$ & $7.33$ & $\ms0.71$ \\
        g & $|02^20\rangle\to|12^20\rangle$ & $1377.14$ & $9.56$ & $0.79$ & $7.44$ & $\ms2.12$ \\
        \hline
        rms &  &  &  &  & $0.108$ &  \\
        \hline
    \end{tabular}
\end{table}

\begin{figure}[h!]
 \begin{center}
    \setlength{\unitlength}{1pt} 
    \includegraphics[width = 8 cm]{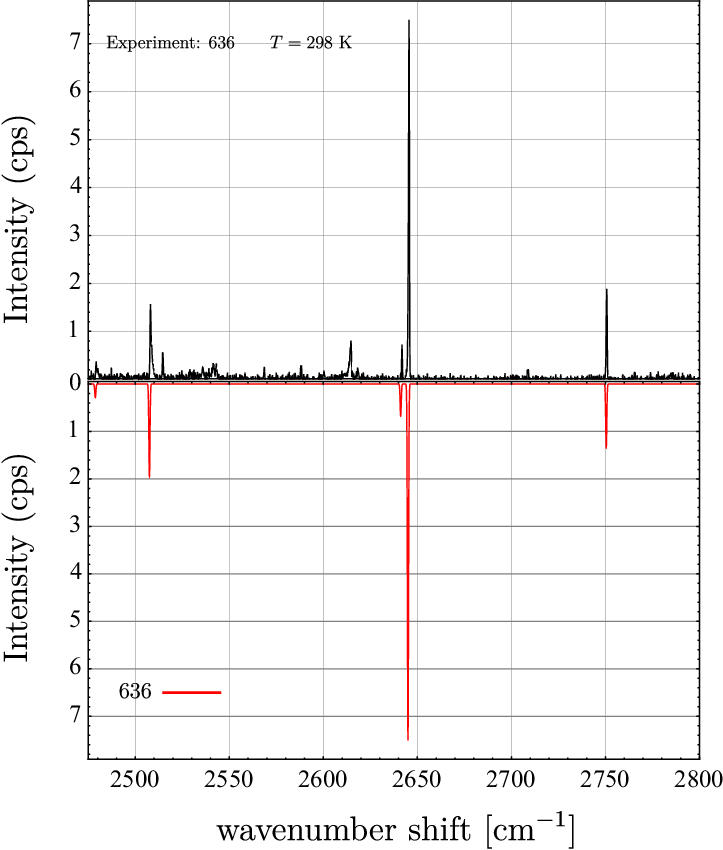}
    \caption{Experimental (top panel, in black) and simulated (bottom panel, in red) Raman spectra for $\mathrm{^{13}C^{16}O_2}$ isotopologue at $T=298.3$~K. The experimental data were taken from Ref.~\cite{Alvarez2026}. Wavenumber shifts were calculated vibrational energies from Ref.~\cite{marisol22}, assuming Gaussian peak profiles with a full width at half maximum (FWHM) of 0.7~cm$^{-1}$.} 
    \label{Figure3}
  \end{center}
\end{figure}

The physical relevant information of a Raman spectrum lies in the relative intensities between spectral lines.  Consequently, once the overall intensity scale is fixed using a single most intense reference line,  the remaining intensities are determined by the transition moments. In Figure~\ref{Figure3} a comparison  shown in mirror image of the experimental spectrum obtained by \'Alvarez {\it et al.} \cite{Alvarez2026} and simulated Raman spectra in the rigid-rotor approximation for pure vibrational transitions ($J=0$) of $^{13}$C$^{16}$O$_2$ at $298.3$ K in the $2470$–$2800$ cm$^{-1}$ spectral range is displayed.   Our simulation shows very good agreement with the experiment, as evidenced by both the relative intensities and the wavenumber shifts $\Delta\bar{\nu}$ calculated using vibrational term values from Refs.~\cite{marisol22,marisol24}. This result confirms that the second $(\partial^2\bar\alpha/\partial \mathcal{S}_g^2)_0$ and third-order derivatives  $(\partial^3\bar\alpha/\partial \mathcal{S}_g\partial\mathcal{S}_+\partial\mathcal{S}_-)_0$  listed in Table~\ref{deriv}, which dominate the transitions in this region, are accurately described by our model.

\begin{figure}[h!]
 \begin{center}
    \setlength{\unitlength}{1pt} 
    \includegraphics[width = \textwidth]{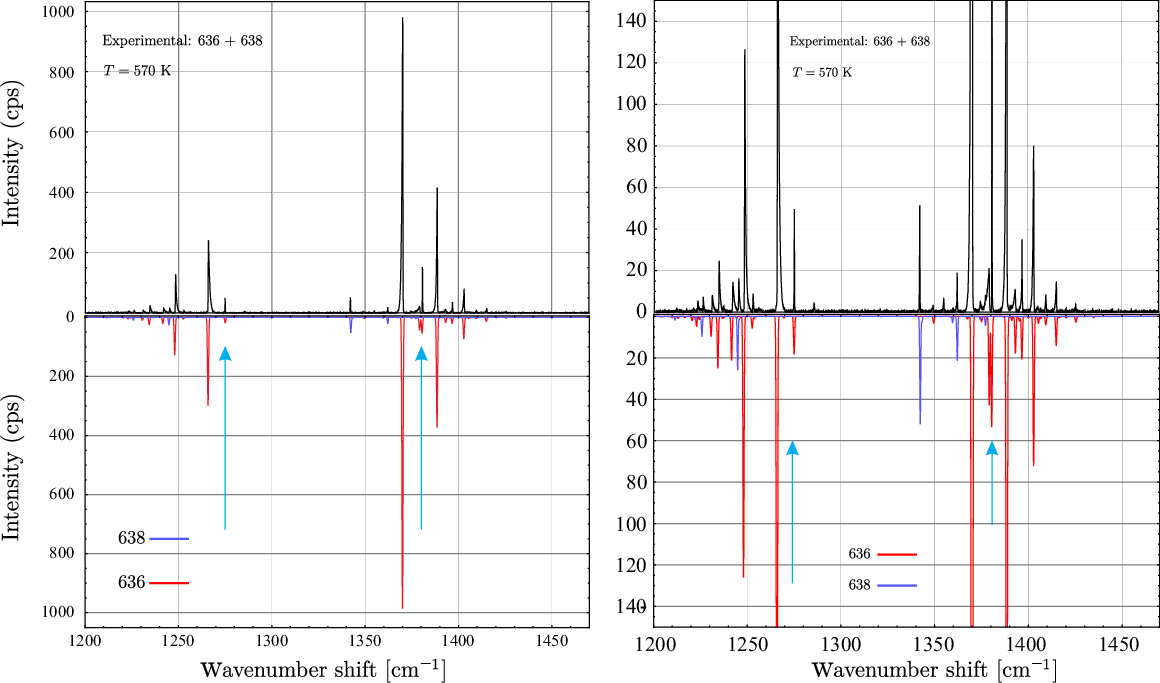}
    \caption{\textcolor{black}{At the left display, the experimental (top panel, in black) and simulated (bottom panel) Raman spectra for $\mathrm{^{13}C^{16}O_2}$ (red lines) and $^{16}$O$^{13}$C$^{18}$O (blue lines) isotopologues at $T=570$~K are compared. At the right a zoom is displayed. The experimental data  were taken from Ref.~\cite{Alvarez2026}. Wavenumber shifts were calculated using vibrational energies from Ref.~\cite{marisol22,marisol24}, assuming Gaussian peak profiles with a FWHM of 0.7~cm$^{-1}$.}} 
    \label{sim570}
  \end{center}
\end{figure}
 
Figure~\ref{sim570} displays, for the mixture of $^{13}$C$^{16}$O$_2$ and $^{16}$O$^{13}$C$^{18}$O isotopologues, the experimental \cite{Alvarez2026} and \textcolor{black}{simulated Raman spectra at $570$~K within the rigid-rotor approximation}. We observe an overall good agreement in both line positions and intensities, albeit  we identify two lines at $1275.03$ and $1380.46$ cm$^{-1}$ with markedly different intensities, as highlighted in the spectrum with blue  arrows. This observation seems to be  in contradiction with the calculated transition moments, marked with an asterisk in Table~\ref{TMET}, since the residuals with respect to the experimental values fall within the range of experimental uncertainty.  In order to elucidate this point, we designed the experiment described in Section~\ref{experiment} to obtain the Raman spectrum of the same isotopologues under comparable temperature conditions.

\begin{figure}[h!]
 \begin{center}
    \setlength{\unitlength}{1pt} 
    \includegraphics[width=\textwidth]{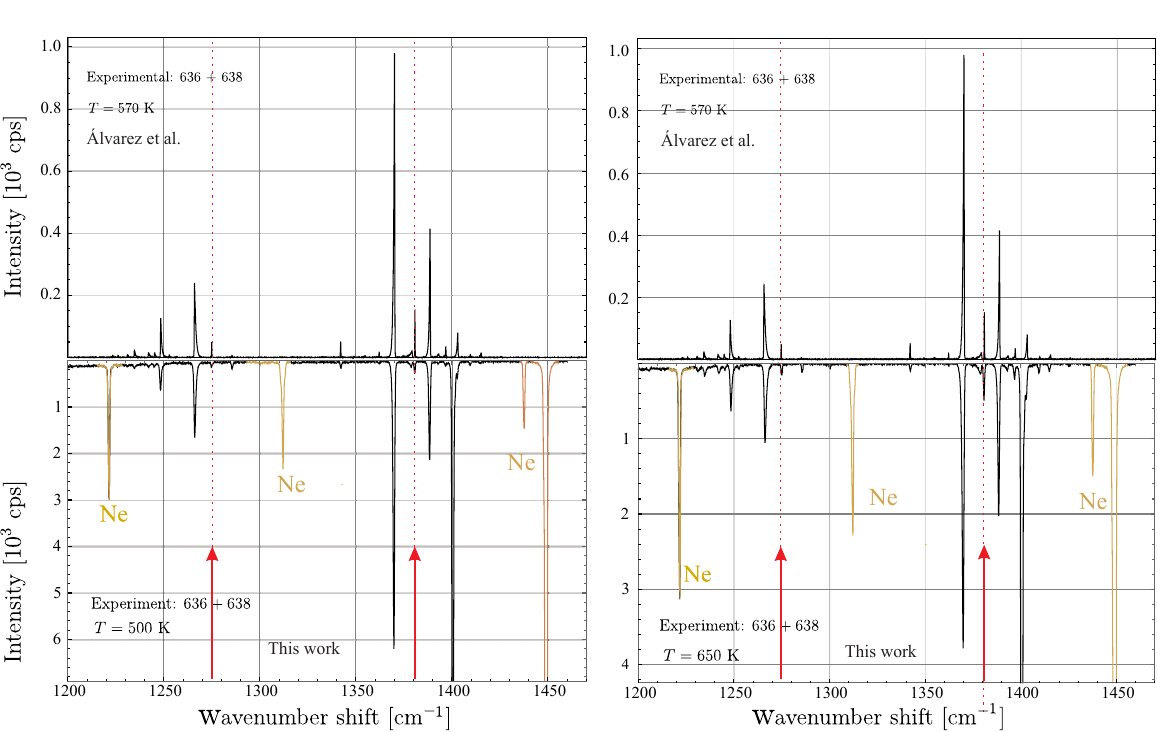}
    \caption{\textcolor{black}{Comparison between the experimental Raman spectrum reported in Ref.~\cite{Alvarez2026} at $T=570$ K (top panel) and the spectrum obtained in this work at $T=500$ K (bottom left) and $T=650$ K (bottom right). Although the spectra were not recorded at the same temperature, they exhibit the same characteristic transitions for both isotopologues.}} 
    \label{Figure4}
  \end{center}
\end{figure}

In Figure~\ref{Figure4}, we present a comparison between the Raman spectra reported by \'Alvarez {\it et al.}~\cite{Alvarez2026} at $570$~K and our experimental results for $500$~K and $650$~K. The two lines discussed above are indicated with arrows.  The experimental spectrum of \'Alvarez {\it et al.}~\cite{Alvarez2026} display the transition lines for the isotopologues $\mathrm{^{13}C^{16}O_2}$ (636) and $^{16}$O$^{13}$C$^{18}$O (638). The lines from Ne are highlighted in color brown. The two arrows indicate the discrepancy observed for the two experimental recordings. In addition, in Figure~\ref{Figure5}, we present the simulated Raman spectra compared with our experimental results. We readily identify our simulations closely fitted to our experimental spectrum. The present results demonstrate that the simulation accurately reproduces the experimentally measured spectrum.

\begin{figure}[h!]
 \begin{center}
    \setlength{\unitlength}{1pt} 
    \includegraphics[width=\textwidth]{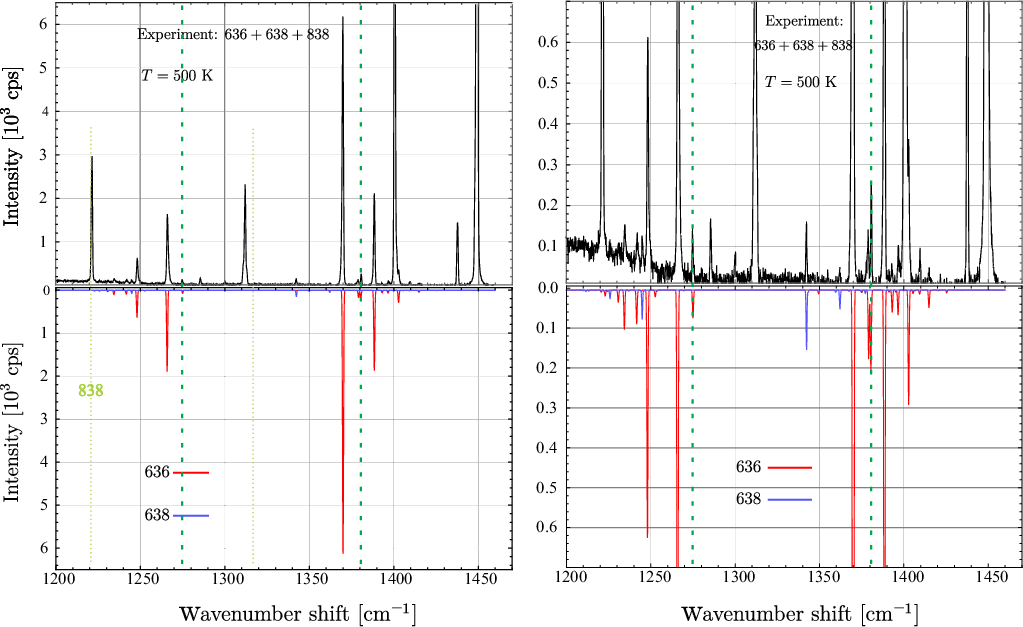}
    \caption{Comparison between the experimental (upper panels, black lines) Raman spectra obtained in this work and the corresponding simulations (bottom panels, colored lines). The lines previously discussed are indicated in green dash lines. At the right a zoom of the spectrum is shown.} 
    \label{Figure5}
  \end{center}
\end{figure}

In order to identify the origin of the discrepancy, we shall pay special attention to the experimental spectrum reported in Ref.~\cite{Alvarez2026} in the vicinity of 1380.46 cm$^{-1}$, as depicted in the left panel of Figure~\ref{Figure6}. Just below the experimental spectrum, our simulated spectrum is shown in mirror image, based on the rigid-rotor approximation and assuming Gaussian peak profiles with a full width at half maximum (FWHM) of 0.7 cm$^{-1}$. Differences in both the intensities and shapes of the lines are evident. We have identified that these discrepancies arise from the rigid-rotor approximation. \textcolor{black}{In the left panel, the simulation is performed considering only the vibrational structure.} In the right panel of Figure~\ref{Figure6}, we present our simulation including the ro-vibrational interaction through Eq.~(\ref{RTI}) by incorporating the rotational diagonal terms into the Raman shifts.
\begin{equation}
\label{rovibener}
\Delta \bar \nu=G_{\nu'}-G_{\nu} +(B_{\nu'}-B_{\nu})\, [J(J+1)-\ell^2],
\end{equation}
where $B_\nu$ are the vibrational-state-dependent rotational constants and $\ell$ the corresponding vibrational angular momentum.
\begin{figure}[h!]
 \begin{center}
    \setlength{\unitlength}{1pt} 
    \includegraphics[width=\textwidth]{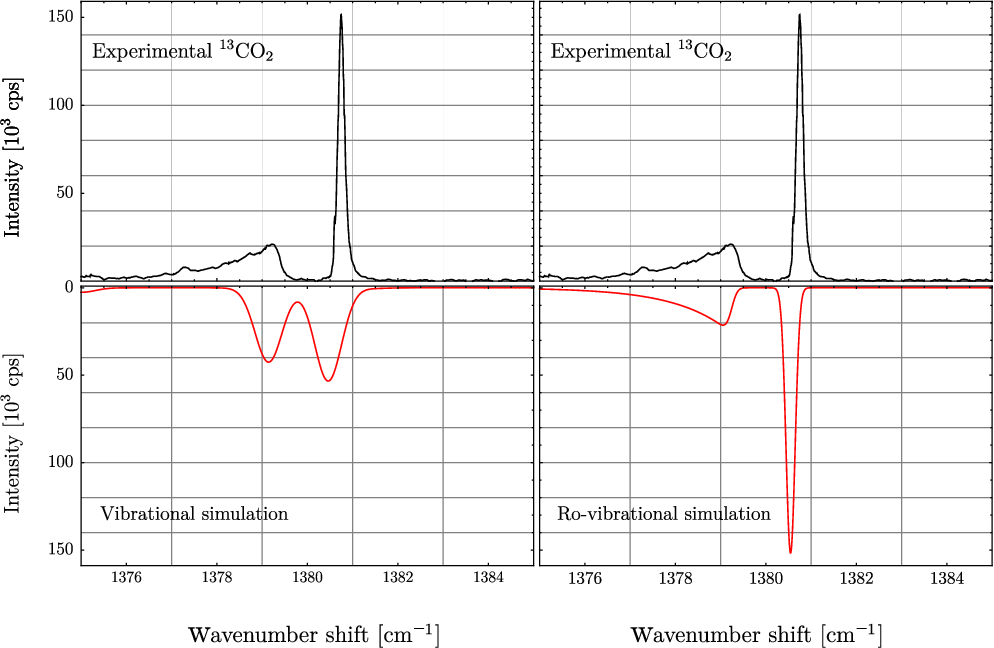}
    \caption{Comparison between the experimental Raman spectrum from Ref.~\cite{Alvarez2026} (top panel, black lines) and the corresponding simulations (bottom panel, red lines). In the vibrational simulation (bottom left), wavenumber shifts were computed using vibrational energies from Ref.~\cite{marisol24}, assuming Gaussian profiles with FWHM of 0.7~cm$^{-1}$. In the ro-vibrational simulation (bottom right), Raman shifts were obtained from Eq.~(\ref{rovibener}) using observed term values from Refs.~\cite{Rothman1992,Miller2004}, with Gaussian profiles of FWHM 0.24~cm$^{-1}$.} 
    \label{Figure6}
  \end{center}
\end{figure}

The correction of the spectrum in both shape and intensity is remarkable. This improvement arises from the rotational contributions to the intensities. In Figure~\ref{stick} we show the ro-vibrational stick spectrum associated with bands~$14$ and~$15$ appearing in Figure~\ref{Figure6} and calculated using Eq.~(\ref{RTI}). Each rotational contribution adds to the total intensity after convolution, which explains the resulting correction in both the shape and magnitude of the spectral features. More precisely, the wider profile of the band on the left is due to the relatively large differences between the rotational constants $B_\nu$ (in cm$^{-1}$) involved in the transition, which fall within the interval $[0.38969,0.39096]$, whereas for the band on the right the interval $[0.38972,0.38973]$ is much narrower. This explains the narrow peak observed on the right.

\begin{figure}[h!]
 \begin{center}
    \setlength{\unitlength}{1pt} 
    \includegraphics[width=0.6\textwidth]{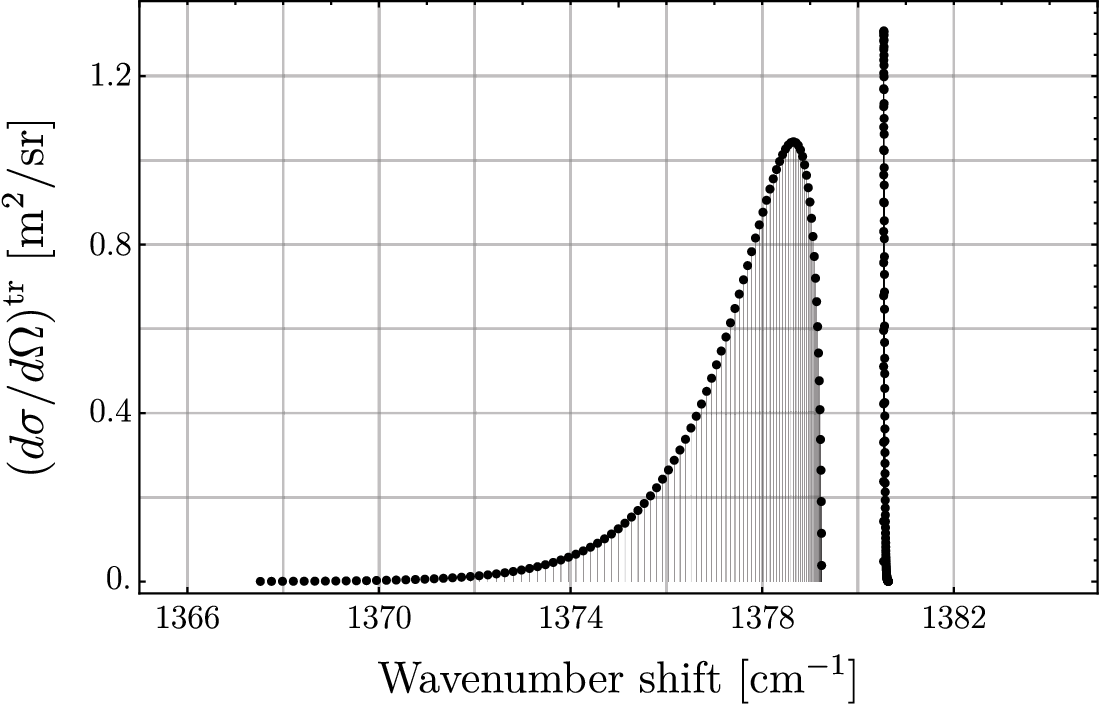}
    \caption{Ro-vibrational stick spectrum of the $^{13}$C$^{16}$O$_2$ isotopologue, with line intensities calculated using Eq.~(\ref{RTI}) with a maximum rotational quantum number of $J_{\rm max}=100$, for bands 14 and 15 listed in Table~\ref{tm636}.
} 
    \label{stick}
  \end{center}
\end{figure}

From this analysis it is clear that the experiment presented by \'Alvarez \textit{et al.} displays information about the ro-vibrational transitions, while our experiment records only the purely vibrational spectra. \textcolor{black}{This can be partially attributed to the experimental resolution, to which the acquisition time of the experiment contributes. Our experiment was measured for $300$~s of time with three accumulations giving information of the most probable events provided by  the vibrational transitions}, \textcolor{black}{whereas the {\'Alvarez} experiment was recorded in 50 minutes \cite{Alvarez2026}, allowing the ro-vibrational transitions to be detected.} This result is quite important when comparing a simulation to the experimental spectrum. In Figure \ref{figure7} we present the simulation of the Raman spectrum previously discussed in Figure \ref{sim570}, but now taking into account (\ref{rovibener}) in the simulation. It is evident the remarkable improvement, as expected from our previous analysis. This study suggests that a Raman simulation    must include the rotational energy structure in order to obtain a faithful reproduction of   the experiment.  However, as noticed, the exact position of the lines are not reproduced. This is explained by the vibrational fitting to the $G_\nu$ term values, %The reason is that we are fitting the vibrational energies  based on the values  $G_\nu$,
which is an approximation to the total experimental energy. Reproduction of  exact values must take into account additional contributions due to the full ro-vibrational interactions.

In Ref.~\cite{Lill1}, when the Raman simulation for the principal isotopologue is compared with that presented in Ref.~\cite{Lemus2014}, the authors suggest that the simulations could be improved by taking into account the anisotropic contribution to the molecular polarizability tensor. However, the present results indicate that the most relevant aspect for improving the spectral description is the incorporation of the rotational energy structure, at least through diagonal corrections to the energy levels. The simulations shown in Figures~\ref{Figure6} and \ref{figure7} suggest that it may not be necessary to include ro-vibrational couplings in the wave functions to generate a high-quality spectrum.  Of course, fine details of the spectrum will be due to the ro-vibrational interactions. Anisotropic contributions to the intensities are also to be present, although it is expected to be smaller than $5$~\% \cite{Alvarez2024}. These contributions can be calculated through an expansion of the polarizability involving the anisotropy, with the corresponding derivatives treated as fitting parameters to experimental transition moments. Regarding these transition moments, no measurements have been reported for carbon dioxide.

\begin{figure}[h!]
 \begin{center}
    \setlength{\unitlength}{1pt} 
    \includegraphics[width=\textwidth]{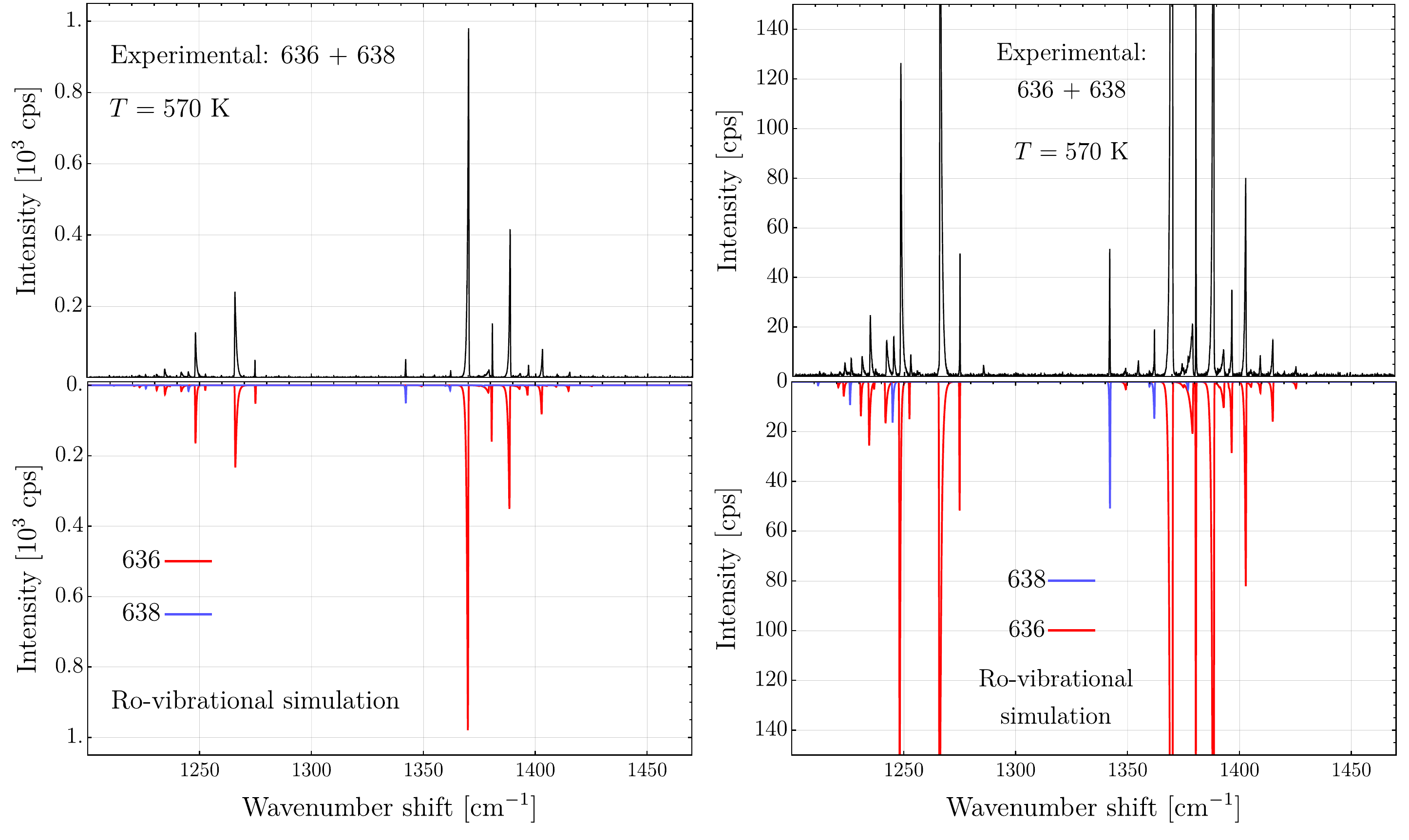}
    \caption{Simulated (bottom panel) and experimental Raman spectra (top panel) for $\mathrm{^{13}C^{16}O_2}$ (red line) and $^{16}$O$^{13}$C$^{18}$O (blue line) isotopologues at $T=570$~K. The experimental data (black line) were taken from Ref.~\cite{Alvarez2026}. Wavenumber shifts were calculated from Eq.~(\ref{rovibener}) using term values from Refs.~\cite{Rothman1992,Miller2004,cerezo,Robert2013}, assuming Gaussian peak profiles with a FWHM of 0.24~cm$^{-1}$.} 
    \label{figure7}
  \end{center}
\end{figure}

\section {Summary and conclusions} 
\label{conclusions}

In this contribution we present the simulation of the Raman spectra of the isotopologues  $^{12}$C$^{16}$O$_2$, $^{13}$C$^{16}$O$_2$ and $^{16}$O$^{13}$C$^{18}$O, with a comparison  to two independent experimental spectra. To achieve this goal we first generate the vibrational wave functions in the framework of the $SU_1(2) \times SU(3) \times SU_2(2)$ model, where the stretching degrees of freedom are mapped to Morse oscillators  and the bending modes are described in terms of the $U(3)$ model. Once the vibrational wave functions are obtained, the derivatives of the polarizability are fitted using $42$ experimental transition moments for the principal isotopologue. Assuming the BO approximation, the estimation of the derivatives for  the principal isotopologue allowed us to calculate the  transition moments for any  isotopologue and their  corresponding  Raman spectra. In particular, we focused our attention to the  spectra of the isotopologues $^{12}$C$^{16}$O$_2$, $^{13}$C$^{16}$O$_2$ and $^{16}$O$^{13}$C$^{18}$O because of the recent experimental Raman spectra recorded at temperatures $298$~K and $570$~K. 
Quite good  descriptions were obtained for the isotopologues $^{12}$C$^{16}$O$_2$ at $1780$~K and $^{13}$C$^{16}$O$_2$ at $298$~K~\cite{Alvarez2024,Alvarez2026}.
With regard to our  simulation of its mixture with $^{16}$O$^{13}$C$^{18}$O, considering the rigid-rotor approximation, two transitions at $1275.03$ cm$^{-1}$ and $1380.46$ cm$^{-1}$ were not appropriately described, with experimental lines  quite large compared to our predictions.
To elucidate the origin of this discrepancy we recorded the experimental Raman spectrum of a mixture of the same  isotopologues at similar temperature conditions. In this latter case the agreement between the experimental and simulated spectra is quite reasonable.

In order to identify the reasons of the discrepancies between both experiments we focused on a small window of the spectrum located in the range $1375-1385$ cm$^{-1}$, where one of the two differing lines appears. Comparing our simulation to the experimental spectrum obtained by {\'Alvarez} \textit{et al.}, large  discrepancies in both the contour and the intensity of the lines are evident. We found out that both differences are due to the lack of the rotational structure. When, in our simulation, the diagonal ro-vibrational corrections are added in the transition energies, both the form and the intensities present a remarkable transformation in accordance to the  experiment. It is important to emphasize that the improvement in the intensities is due to the diagonal corrections to the energy, but not in the wave functions, which continue to be in the rigid-rotor approximation. This result allows us to conclude that the features of the experiment establishes the possibility  to be described either with a pure vibrational spectrum or taking rotational effects in the energy. Even taking into account the latter effects, the simulations fails in reproducing the exact position of the lines. This is because we are considering the  energies as a sum of vibrational, rotational and ro-vibrational contributions, keeping in our fitting approach only the vibrational terms. Therefore, for an approach including the ro-vibrational interactions, the lines will be fitted at the experimental positions.  On the basis of this analysis we conclude that anisotropic effects are not so important as suggested in Ref.~\cite{Lill1}.

A possible improvement to our description  may be the introduction in our formalism the polyad mixing in the wave functions and, consequently, in the Raman spectra. Regarding these effects, resonances at  upper  energy regions  should be explored to identify the significant ones. Such resonances are being analyzed and chosen appropriately to be taken into account as perturbations. This work is in progress.

\section*{Acknowledgments}
The authors gratefully acknowledge Carlos \'Alvarez, Guzm\'an Tejeda and Jos\'e M. Fern\'andez for kindly providing some of the experimental data used in this work.

\section*{Funding}
This work is partially supported by DGAPA-UNAM, México, under project IN-212224, and  project BiMiAb\_H2 at the Federal Institute for Geosciences and Natural Resources, Germany. This project has also received funding from the European Union's Horizon 2020 research and innovation program under Marie Sklodowska-Curie grant agreement No. 872081, and grant PID2022-136228NB-C21 (M.C.) funded by MCIN/AEI/10.13039/501100011033, and, as appropriate, by "ERDF A way of making Europe", the "European Union", or the "European Union NextGenerationEU/PRTR". This work is also supported by the Consejería de Transformación Económica, Industria, Conocimiento y Universidades, Junta de Andalucía and European Regional Development Fund (ERDF 2021-2027) under the project EPIT1462023 (M.C.).

\bibliographystyle{elsarticle-num}
%\nocite{*}
\bibliography{ref}

\end{document}